\documentclass[english]{revtex4}
\usepackage[T1]{fontenc}
\usepackage[latin9]{inputenc}
\usepackage{amsmath}
\usepackage{amssymb}

\makeatletter
\@ifundefined{textcolor}{}
{%
 \definecolor{BLACK}{gray}{0}
 \definecolor{WHITE}{gray}{1}
 \definecolor{RED}{rgb}{1,0,0}
 \definecolor{GREEN}{rgb}{0,1,0}
 \definecolor{BLUE}{rgb}{0,0,1}
 \definecolor{CYAN}{cmyk}{1,0,0,0}
 \definecolor{MAGENTA}{cmyk}{0,1,0,0}
 \definecolor{YELLOW}{cmyk}{0,0,1,0}
}

\usepackage{babel}

\makeatother

\usepackage{babel}
\begin{document}
\title{Unitary, continuum, stationary perturbation theory for the radial
Schrödinger equation}
\author{Scott E. Hoffmann}
\address{School of Mathematics and Physics,~\\
 The University of Queensland,~\\
 Brisbane, QLD 4072~\\
 Australia}
\email{scott.hoffmann@uqconnect.edu.au}

\begin{abstract}
The commutators of the Poincaré group generators will be unchanged
in form if a unitary transformation relates the free generators to
the generators of an interacting relativistic theory. We test the
concept of unitary transformations of generators in the nonrelativistic
case, requiring that the free and interacting Hamiltonians be related
by a unitary transformation. Other authors have applied this concept
to time-dependent perturbation theory to give unitarity of the time
evolution operator to each order in perturbation theory, with results
that show improvement over the standard perturbation theory. In our
case, a stationary perturbation theory can be constructed to find
approximate solutions of the radial Schrödinger equation for scattering
from a spherically symmetric potential. General formulae are obtained
for the phase shifts at first and second order in the coupling constant.
We test the method on a simple system with a known exact solution
and find complete agreement between our first- and second-order contributions
to the $s$-wave phase shifts and the corresponding expansion to second
order of the exact solution.
\end{abstract}
\maketitle

\section{Introduction}

The motivation for this paper came from general considerations of
interacting relativistic, quantum-mechanical theories. The proposal
arising from those considerations was that the Poincaré generators
of the interacting theory be related to those of the free theory by
a unitary transformation. In this paper, we test that proposal on
a nonrelativistic system, that of scattering of one spinless particle
from a spherically symmetric potential. There, the interacting Hamiltonian
is to be the unitary transformation of the free Hamiltonian. We find
that a perturbation theory can be constructed by expanding the generator
of the unitary transformation in powers of the coupling constant.
This perturbation theory, because of the unitarity of the transformation
acting on the state vectors, has the property that normalization is
unchanged at each order in the coupling constant. Such is not the
case for the conventional Green function method, with which we compare
in section \ref{sec:Comparison-with-the}.

Casas \textit{et al}. \citep{Casas2011} considered time-dependent
perturbation theory, but their unitary transformation result carries
over to the stationary perturbation theory considered here. They proposed
that a unitary transformation of the original Hamiltonian be constructed
to make an alternative Hamiltonian that is easier to solve
\[
H_{i}^{\prime}=T^{\dagger}H_{i}T,
\]
with $T,$ unitary, to be determined. The simplest choice is making
$H^{\prime}$ equal to $H_{f},$ which we assume can be solved exactly.
This case is then the one considered here, with
\[
T=U_{if}
\]
in our notation, a unitary transformation that depends on the coupling
constant. In their examples it can also depend explicitly on time.
Then the unitarity of their time evolution operator,
\[
U(t)=T\,e^{-iH_{f}t},
\]
is guaranteed at every order in their perturbation expansion. Their
tests of the method on two-level systems show improvements over the
standard time-dependent perturbation theory.

Ali \citep{Ali2000} also considered time-dependent perturbation theory
and the importance of unitarity. He noted that the Born series \citep{Born1926}
for the time evolution operator is not unitary to any finite order.
He proposed an exponentiation of the terms to a finite order, with
modifications necessary to guarantee unitarity. For example, at first
order he takes
\[
U^{(1)}(t)\rightarrow\exp(-i\int_{0}^{t}dt^{\prime}\,H_{I}(t^{\prime}))
\]
instead of
\[
U_{\mathrm{Born}}^{(1)}(t)=1-i\int_{0}^{t}dt^{\prime}\,H_{I}(t^{\prime}),
\]
which is only unitary to first order. His tests of the method on two-level
systems show improvements over the standard time-dependent perturbation
theory.

For a free, special-relativistic, quantum-mechanical theory, we must
have representations of the Poincare generators $P_{f}^{\mu}$ (the
four components of total energy-momentum which are the generators
of spacetime translations) and $M_{f}^{\mu\nu}$ (with $K^{i}=M^{0i},$
the boost generators and $J^{k}=\frac{1}{2}\epsilon_{kij}M^{ij},$
the angular momenta, generators of rotations) satisfying the commutation
relations
\begin{align}
[P_{f}^{\mu},P_{f}^{\nu}] & =0,\nonumber \\{}
[M_{f}^{\mu\nu},P_{f}^{\lambda}] & =i(g^{\nu\lambda}P_{f}^{\mu}-g^{\mu\lambda}P_{f}^{\nu}),\nonumber \\{}
[M_{f}^{\mu\nu},M_{f}^{\rho\sigma}] & =i(g^{\mu\sigma}M_{f}^{\nu\rho}+g^{\nu\rho}M_{f}^{\mu\sigma}-g^{\mu\rho}M_{f}^{\nu\sigma}-g^{\nu\sigma}M_{f}^{\mu\rho}).\label{eq:1}
\end{align}
For an interacting theory we must have different generators, $P_{i}^{\mu}$
and $M_{i}^{\mu\nu},$ satisfying commutators of the same form
\begin{align}
[P_{i}^{\mu},P_{i}^{\nu}] & =0,\nonumber \\{}
[M_{i}^{\mu\nu},P_{i}^{\lambda}] & =i(g^{\nu\lambda}P_{i}^{\mu}-g^{\mu\lambda}P_{i}^{\nu}),\nonumber \\{}
[M_{i}^{\mu\nu},M_{i}^{\rho\sigma}] & =i(g^{\mu\sigma}M_{i}^{\nu\rho}+g^{\nu\rho}M_{i}^{\mu\sigma}-g^{\mu\rho}M_{i}^{\nu\sigma}-g^{\nu\sigma}M_{i}^{\mu\rho}).\label{eq:2}
\end{align}
We note that the form of a commutator is invariant under a unitary
transformation. If
\begin{equation}
[A,B]=iC\label{eq:3}
\end{equation}
and
\begin{equation}
A^{\prime}=UAU^{\dagger},\quad B^{\prime}=UBU^{\dagger},\quad C^{\prime}=UCU^{\dagger},\label{eq:4}
\end{equation}
then
\begin{align}
[A^{\prime},B^{\prime}] & =[UAU^{\dagger},UBU^{\dagger}]=U[A,B]U^{\dagger}=iUCU^{\dagger}=iC^{\prime}.\label{eq:5}
\end{align}

So we propose that the free generators and interacting generators
be related by a unitary transformation
\begin{equation}
P_{i}^{\mu}=U_{if}P_{f}^{\mu}U_{if}^{\dagger}\quad\mathrm{and}\quad M_{i}^{\mu\nu}=U_{if}M_{f}^{\mu\nu}U_{if}^{\dagger}.\label{eq:6}
\end{equation}
This unitary transformation will contain all the information on the
interaction, so must depend on the charge. Also state vectors will
be related by
\begin{equation}
|\,\psi_{i}\,\rangle=U_{if}\,|\,\psi_{f}\,\rangle,\label{eq:7}
\end{equation}
where $|\,\psi_{f}\,\rangle$ is a free state vector and $|\,\psi_{i}\,\rangle$
is the corresponding interacting state vector. The consequences of
this proposal will be considered in a future work.

We test the proposal on a nonrelativistic problem, the scattering
of a spinless particle from a spherically symmetric potential. There,
in the free theory, the first of the commutators
\begin{equation}
[\boldsymbol{J}_{f},H_{f}]=0\quad\mathrm{and}\quad[J_{f}^{i},J_{f}^{j}]=i\epsilon_{ijk}J_{f}^{k}\label{eq:8}
\end{equation}
expresses the rotational invariance of the Hamiltonian. In the interacting
theory, we require commutators of the same form,
\begin{equation}
[\boldsymbol{J}_{i},H_{i}]=0\quad\mathrm{and}\quad[J_{i}^{i},J_{i}^{j}]=i\epsilon_{ijk}J_{i}^{k}.\label{eq:9}
\end{equation}
So we propose a unitary connection between the free operators and
the interacting operators,
\begin{equation}
H_{i}=U_{if}\,H_{f}\,U_{if}^{\dagger}\quad\mathrm{and}\quad\boldsymbol{J}_{f}=U_{if}\,\boldsymbol{J}_{f}\,U_{if}^{\dagger}.\label{eq:10}
\end{equation}
We must also be able to write the interacting Hamiltonian in terms
of a potential,
\begin{equation}
H_{i}=H_{f}+V\equiv H_{f}+\lambda\,U,\label{eq:11}
\end{equation}
where $\lambda$ is a dimensionless coupling constant. Note $U$ is
Hermitian but not unitary.

As we will see below, these equations can be solved for the generator
of $U_{if}$ as a series in powers of $\lambda,$ giving a perturbation
theory that differs from those previously considered. In particular,
once $U_{if}$ is obtained to the desired order, the unitary transformation
of the state vectors, for example the energy eigenvectors, is
\begin{equation}
|\,E;i\,\rangle=U_{if}\,|\,E;f\,\rangle.\label{eq:12}
\end{equation}
Normalization will be preserved to each order, with no need for renormalization.

The organization of this paper is as follows. In section \ref{sec:Unitary-transformation-for}
we develop the unitary transformation theory for a spherically symmetric
potential in nonrelativistic quantum mechanics. The result will be
expressions for the unitary transformation generator at first and
second order in $\lambda.$ In section \ref{sec:The-spherical-well}
we consider a particular model, the spherical well or barrier, for
which an exact solution is known. We will find agreement with the
exact solution for the phase shifts to order $\lambda^{2}.$ In section
\ref{sec:The-Wronskian-formula}, we find general forms for the first
and second order phase shifts using the phase shift formula derived
using the Wronskian theorem. We then compare our results with the
conventional Green function method, which is not designed to preserve
normalization, in section \ref{sec:Comparison-with-the}. Conclusions
follow in section \ref{sec:Conclusions}.

\section{\label{sec:Unitary-transformation-for}Unitary transformation for
spherically symmetric potentials}

Our aim is to solve the radial Schrodinger equation \citep{Weinberg2015}
\begin{equation}
\{-\frac{1}{2m}\frac{d^{2}}{dr^{2}}+\frac{l(l+1)}{2mr^{2}}+V(r)\}y_{l}^{(i)}(r,p)=\frac{p^{2}}{2m}y_{l}^{(i)}(r,p)\label{eq:2.1}
\end{equation}
on $r\geq0,$ with boundary condition $y_{l}^{(i)}(0,k)=0.$ We write
\begin{equation}
V(r)=\lambda\,U(r).\label{eq:2.2}
\end{equation}
We take as the free Hamiltonian
\begin{equation}
H_{f}=-\frac{1}{2m}\frac{d^{2}}{dr^{2}}+\frac{l(l+1)}{2mr^{2}}.\label{eq:2.3}
\end{equation}
The free solutions that vanish at the origin are
\begin{equation}
y_{l}^{(f)}(r,p)=\sqrt{\frac{2}{\pi}}\,kr\,j_{l}(pr),\label{eq:2.4}
\end{equation}
normalized to
\begin{equation}
\int_{0}^{\infty}dr\,y_{l}^{(f)}(r,k_{1})\,y_{l}^{(f)}(r,k_{2})=\delta(k_{1}-k_{2}).\label{eq:2.5}
\end{equation}
There are also solutions singular at the origin (for $l\geq1$)
\begin{equation}
\tilde{y}_{l}^{(f)}(r,p)=\sqrt{\frac{2}{\pi}}\,pr\,n_{l}(pr),\label{eq:2.6}
\end{equation}
which we will require in section \ref{sec:Comparison-with-the}.

We write the unitary transformation in terms of an Hermitian generator,
$\Theta,$ as
\begin{equation}
U_{if}=e^{-i\Theta}.\label{eq:2.7}
\end{equation}
We write $\Theta$ as a series in powers of $\lambda,$ noting that
it must vanish for $\lambda=0$ to return the free theory:
\begin{equation}
\Theta=\lambda\,\Theta^{(1)}+\frac{1}{2}\lambda^{2}\Theta^{(2)}+\dots.\label{eq:2.8}
\end{equation}
Then
\begin{equation}
U_{if}=1-i\lambda\,\Theta^{(1)}-\frac{i}{2}\lambda^{2}\Theta^{(2)}-\frac{1}{2}\lambda^{2}\Theta^{(1)2}+\dots\label{eq:2.9}
\end{equation}
to this order.

Then from the two representations of $H_{i}$ in Eqs. (\ref{eq:10})
and (\ref{eq:11}), we require
\begin{align}
U_{if}\,H_{f}\,U_{if}^{\dagger} & =H_{f}+\lambda\,U.\label{eq:2.10}
\end{align}
To $\mathcal{O}(\lambda^{2}),$ this is
\begin{equation}
H_{f}+i\lambda[H_{f},\Theta^{(1)}]+\frac{i\lambda^{2}}{2}[H_{f},\Theta^{(2)}]-\frac{1}{2}\lambda^{2}[\Theta^{(1)},[\Theta^{(1)},H_{f}]]=H_{f}+\lambda\,U.\label{eq:2.11}
\end{equation}
Equating like powers of $\lambda,$ this gives
\begin{equation}
i[H_{f},\Theta^{(1)}]=U\quad\mathrm{and}\quad[H_{f},\Theta^{(2)}]=[\Theta^{(1)},U].\label{eq:2.12}
\end{equation}

Taking matrix elements in the free basis, with
\begin{equation}
\Theta_{l}^{(n)}(k_{1},k_{2})\equiv\langle\,k_{1},l\,|\,\Theta^{(n)}\,|\,k_{2},l\,\rangle\quad\mathrm{for}\ n=1,2,3,\dots\quad\mathrm{and}\quad U_{l}(k_{1},k_{2})\equiv\langle\,k_{1},l\,|\,U\,|\,k_{2},l\,\rangle,\label{eq:2.13}
\end{equation}
gives the solutions
\begin{equation}
\Theta_{l}^{(1)}(k_{1},k_{2})=-i\frac{2m\,U_{l}(k_{1},k_{2})}{k_{1}^{2}-k_{2}^{2}}\label{eq:2.14}
\end{equation}
and
\begin{align}
\Theta_{l}^{(2)}(k_{1},k_{2}) & =\frac{2m}{k_{1}^{2}-k_{2}^{2}}\int_{0}^{\infty}dk\,\{\Theta_{l}^{(1)}(k_{1},k)U_{l}(k,k_{2})-U_{l}(k_{1},k)\Theta_{l}^{(1)}(k,k_{2})\}\nonumber \\
 & =-i\frac{2m}{k_{1}^{2}-k_{2}^{2}}\int_{0}^{\infty}dk\,\{\frac{2m\,U_{l}(k_{1},k)U_{l}(k,k_{2})}{k_{1}^{2}-k^{2}}-\frac{2m\,U_{l}(k_{1},k)U_{l}(k,k_{2})}{k^{2}-k_{2}^{2}}\}.\label{eq:2.15}
\end{align}

Note that from the Hermiticity of $U,$ $U_{l}(k_{1},k_{2})$ is always
symmetric in $k_{1}$ and $k_{2}.$ In general (and in the example
we consider in section \ref{sec:The-spherical-well}) it will not
vanish at $k_{1}=k_{2}.$ So the expression in Eq. (\ref{eq:2.14})
is singular at $k_{1}=k_{2}.$ We deal with this singularity by imposing
a rule of principal part integration. The justification for this is
that, at first order, it ensures that the correction state vector
is orthogonal to the unperturbed state vector, as required for unit
normalization to first order. This leads to finite results in agreement
with previous calculations.

We note the similarity of our result to the expression from first
order perturbation theory of a discrete spectrum:
\begin{align*}
|\,n\,(1)\,\rangle & =|\,n\,(0)\,\rangle-\lambda\sum_{n^{\prime}\neq n}|\,n^{\prime}\,(0)\,\rangle\,\frac{\langle\,n^{\prime}\,(0)\,|\,V\,|\,n\,(0)\,\rangle}{E_{n^{\prime}}^{(0)}-E_{n}^{(0)}},
\end{align*}
where the absence of the contribution with $n^{\prime}=n$ gives a
finite result and unitarity of the transformation to first order.
We can consider our method to be the continuum limit of the discrete
method.

Then the unitarily transformed solution, to $\mathcal{O}(\lambda)$,
for momentum eigenvalue $p,$ is
\begin{align}
\langle\,r\,|\,p,l;i\,(1)\,\rangle & =\int_{0}^{\infty}dk\,\langle\,r\,|\,k,l;f\,\rangle\langle\,k,l;f\,|\,U_{if}\,|\,p,l;f\,\rangle,\nonumber \\
y_{l}^{(i,1)}(r,p) & =y_{l}^{(f)}(r,p)-\lambda\,P\int_{0}^{\infty}dk\,y_{l}^{(f)}(r,k)\frac{2m\,U_{l}(k,p)}{(k^{2}-p^{2})}.\label{eq:2.16}
\end{align}

There are two ways to extract the phase shifts from this expression.
The first is to evaluate the integral, then find the asymptotic behaviour
of $y_{l}^{(i,1)}(r,p)$ as $r\rightarrow\infty.$ We will do this
to $\mathcal{O}(\lambda^{2})$ for a simple example potential in section
\ref{sec:The-spherical-well} as a test of our method. The other method
is to use the result derived using the Wronskian theorem \citep{Weinberg2015}
\begin{equation}
\sin\delta_{l}(p)=-\frac{\pi m}{p}\,\int_{0}^{\infty}dr\,y_{l}^{(i)}(r,p)V(r)y_{l}^{(f)}(r,p).\label{eq:2.17}
\end{equation}
To use this formula to find the phase shifts correct to $\mathcal{O}(\lambda^{2})$
only requires the interacting wavefunction to $\mathcal{O}(\lambda).$
We will consider this method in section \ref{sec:The-Wronskian-formula}.

\section{\label{sec:The-spherical-well}The spherical well or barrier}

In this section we test the unitary continuum perturbation method
on a simple example, the spherical well or barrier. The potential
is defined as
\begin{equation}
V(r)=\begin{cases}
\frac{\lambda}{R} & 0\leq r\leq R,\\
0 & r>R.
\end{cases}\label{eq:3.1}
\end{equation}
For simplicity, we only consider the case $l=0,$ where the pair of
free fundamental solutions is
\begin{equation}
y_{0}^{(f)}(r,p)=\sqrt{\frac{2}{\pi}}\,\sin(pr),\quad\tilde{y}_{0}^{(f)}(r,p)=-\sqrt{\frac{2}{\pi}}\,\cos(pr).\label{eq:3.2}
\end{equation}

We find that the matrix elements of the potential are
\begin{equation}
\langle\,k_{1},0;f\,|\,V\,|\,k_{2},0;f\,\rangle=\lambda\,U_{0}(k_{1},k_{2})=\frac{\lambda}{\pi}\,\{\mathrm{sinc}((k_{1}-k_{2})R)-\mathrm{sinc}((k_{1}+k_{2})R)\}.\label{eq:3.3}
\end{equation}
Our method gives the normalized solution to $\mathcal{O}(\lambda)$
\begin{equation}
y_{0}^{(i,1)}(r,p)=\sqrt{\frac{2}{\pi}}\,\sin(pr)-\lambda\,P\int_{0}^{\infty}dk\,\sqrt{\frac{2}{\pi}}\,\sin(kr)\frac{2m\,U_{0}(k,p)}{(k^{2}-p^{2})}.\label{eq:3.4}
\end{equation}

With $q=k-p,$ the principal part integral becomes
\begin{equation}
P\int_{0}^{\infty}dk\,\sqrt{\frac{2}{\pi}}\,\sin(kr)\frac{2m\,U_{l}(k,p)}{(k^{2}-p^{2})}=P\int_{-p}^{p}dq\,\sqrt{\frac{2}{\pi}}\,\sin((p+q)r)\frac{2m\,U_{l}(p+q,p)}{(2p+q)q}+\int_{2p}^{\infty}dk\,\sqrt{\frac{2}{\pi}}\,\sin(kr)\frac{2m\,U_{l}(k,p)}{(k^{2}-p^{2})}.\label{eq:3.5}
\end{equation}
We are only interested in the asymptotic behaviour of these integrals
as $r\rightarrow\infty,$ to obtain the phase shifts. The second integral
will vanish like $1/r$ since the integrand is analytic on this region
and the factor $\sin(kr)$ oscillates rapidly with $k.$ In the principal
part integration, the integrand is separated into parts odd in $q$
and even in $q.$ The integral of the odd part on the symmetric interval
$q\in[-p,p]$, even if it is singular at $q=0,$ will vanish. The
principal part integral is defined as the integral from $-p$ to $-\epsilon$
plus the integral from $+\epsilon$ to $+p$ for $\epsilon\ll p,$
with the limit as $\epsilon\rightarrow0^{+}$ taken of the result.

This gives
\begin{multline}
\quad P\int_{-p}^{p}dq\,\sqrt{\frac{2}{\pi}}\,\sin((p+q)r)\frac{2m\,U_{l}(p+q,p)}{(2p+q)q}\\
=\sqrt{\frac{2}{\pi}}\,\sin(pr)\int_{-p}^{p}dq\,\cos(qr)\left\{ \frac{4mp[U_{l}(p+q,p)]_{-}}{(4p^{2}-q^{2})q}-\frac{2m\,[U_{l}(p+q,p)]_{+}}{(4p^{2}-q^{2})}\right\} \\
+\sqrt{\frac{2}{\pi}}\,\cos(pr)\int_{-p}^{p}dq\,\frac{\sin(qr)}{q}\left\{ \frac{4mp[U_{l}(p+q,p)]_{+}}{(4p^{2}-q^{2})}-q\frac{2m[U_{l}(p+q,p)]_{-}}{(4p^{2}-q^{2})}\right\} ,\label{eq:3.6}
\end{multline}
where
\begin{equation}
[U_{l}(p+q,p)]_{\pm}=\frac{1}{2}\{U_{l}(p+q,p)\pm U_{l}(p-q,p)\}\label{eq:3.7}
\end{equation}
are the parts even (+) and odd (-) in $q.$ The first term will vanish
like $1/r$ as $r\rightarrow\infty$ since the integrand is without
singularities and $\cos(qr)$ oscillates rapidly as a function of
$q.$ Then we note
\begin{equation}
\frac{\sin(qr)}{q}=r\,\mathrm{sinc}(qr)\sim\pi\,\delta(q).\label{eq:3.8}
\end{equation}
This factor approaches a delta function in $q$ as $r\rightarrow\infty.$
This will give the only nonvanishing contribution as $r\rightarrow\infty$
\begin{multline}
y_{0}^{(i,1)}(r,p)\rightarrow\sqrt{\frac{2}{\pi}}\,\sin(pr)-\lambda\,\sqrt{\frac{2}{\pi}}\,\cos(pr)\int_{-p}^{p}dq\,\pi\,\delta(q)\frac{2m\,U_{0}(p+q,p)}{2p+q}\\
=\sqrt{\frac{2}{\pi}}\,\sin(pr)+\{-\frac{\pi\lambda}{p/m}U_{0}(p,p)\}\sqrt{\frac{2}{\pi}}\,\cos(pr).\label{eq:3.9}
\end{multline}

This is of the form, to $\mathcal{O}(\lambda),$
\begin{equation}
y_{0}^{(i,1)}(r,p)\rightarrow\sqrt{\frac{2}{\pi}}\,\sin(pr+\delta_{0}^{(1)}(p)),\label{eq:3.10}
\end{equation}
with phase shift
\begin{equation}
\delta_{0}^{(1)}(p)=-\frac{\pi\lambda}{p/m}U_{0}(p,p)=-\frac{\lambda}{p/m}(1-\mathrm{sinc}(2pR)).\label{eq:3.11}
\end{equation}

We note that the size of the phase shift at high energies ($pR\gg1$)
is controlled by the factor
\begin{equation}
\eta\equiv\frac{\lambda}{p/m}.\label{eq:3.12}
\end{equation}
This is very similar to the case of the Coulomb potential, where
\begin{equation}
\eta_{C}=\frac{Z_{1}Z_{2}\,\alpha}{p/m}\label{eq:3.13}
\end{equation}
controls the size of the phase shifts. Here $Z_{1}$ and $Z_{2}$
are the atomic numbers of the target and projectile, respectively
and $\alpha\cong1/137$ is the fine structure constant.

At second order, using Eqs. (\ref{eq:2.14},\ref{eq:2.15}), we find
the contribution
\begin{multline}
\langle\,r\,|\,-\frac{1}{2}\lambda^{2}(\Theta^{(1)2}+i\Theta^{(2)})\,|\,p,0;f\,\rangle\\
=P\int_{0}^{\infty}dk\,\sqrt{\frac{2}{\pi}}\,\sin(kr)\,4m^{2}\lambda^{2}\int_{0}^{\infty}dk^{\prime}\,\frac{U_{0}(k,k^{\prime})}{(k^{2}-k^{\prime2})}U_{0}(k^{\prime},p)\left\{ \frac{1}{(k^{\prime2}-p^{2})}-\frac{1}{(k^{2}-p^{2})}\right\} .\label{eq:3.14}
\end{multline}
We perform the $k$ integral first. In the first term, there is only
a pole at $k=k^{\prime},$ which we treat similarly to the first order
calculation just given. In the second term, there are poles at $k=k^{\prime}$
and $k=p.$ We use expansions of $\sin(kr)$ around each of these
points separately. To proceed with the remaining $k^{\prime}$ integral,
we choose to consider only the regime $pR\gg1$ and obtain results
to order $1/pR.$ We encounter factors
\begin{equation}
\frac{\sin(qR)}{q}\sim\pi\,\delta(q)\quad\mathrm{and}\quad\mathrm{sinc^{2}}(qR)\sim\frac{\pi}{R}\,\delta(q),\label{eq:3.14.1}
\end{equation}
approximations to delta functions in this regime. The net result is
\begin{multline}
\langle\,r\,|\,-\frac{1}{2}\lambda^{2}(\Theta^{(1)2}+i\Theta^{(2)})\,|\,p,0;f\,\rangle\rightarrow\\
\sqrt{\frac{2}{\pi}}\,\sin(pr)\{-\frac{1}{2}\eta^{2}(1-2\mathrm{sinc}(2pR))\}+\sqrt{\frac{2}{\pi}}\,\cos(pr)\{-\eta^{2}\frac{(1+2\cos(2pR))}{2pR}\}+\mathcal{O}(\frac{1}{(pR)^{2}}).\label{eq:3.15}
\end{multline}
The factor in the first term is
\begin{equation}
-\frac{1}{2}\eta^{2}(1-2\mathrm{sinc}(2pR))=-\frac{1}{2}\delta_{0}^{(1)}(p)^{2}+\mathcal{O}(\frac{1}{(pR)^{2}}),\label{eq:3.16}
\end{equation}
as required for unitarity. The second term gives the prediction
\begin{equation}
\delta_{0}^{(2)}(p)=-\eta^{2}\frac{(1+2\cos(2pR))}{2pR}+\mathcal{O}(\frac{1}{(pR)^{2}}).\label{eq:3.17}
\end{equation}

The spherical well and barrier problems can be solved exactly with
elementary methods. The energy shift on $0\leq r\leq R$ gives free
solutions with momenta
\begin{equation}
p^{\prime}=\sqrt{p^{2}+\frac{2m\lambda}{R}}=p\sqrt{1+\frac{2\eta}{pR}}.\label{eq:3.18}
\end{equation}
To satisfy the boundary condition, the solution must be proportional
to $y_{0}^{(f)}(r,p^{\prime})$ in that region. For $r>R,$ the solution
is a linear combination of the fundamental solutions $y_{0}^{(f)}(r,p)$
and $\tilde{y}_{0}^{(f)}(r,p).$ Requiring the wavefunction and its
first derivative to be continuous across the boundary gives the solution,
which then only needs normalization.

We find that the $l=0$ phase shifts, $\delta_{0}(p),$ are given
by
\begin{equation}
e^{i2\delta_{0}(p)}=\frac{(A_{0}-iB_{0})^{2}}{A_{0}^{2}+B_{0}^{2}},\label{eq:3.19}
\end{equation}
with
\begin{align}
A_{0} & =\kappa^{2}j_{0}(\kappa^{\prime})n_{0}^{\prime}(\kappa)-\kappa^{\prime}\kappa\,n_{0}(\kappa)j_{0}^{\prime}(\kappa^{\prime}),\nonumber \\
B_{0} & =\kappa^{\prime}\kappa\,j_{0}(\kappa)j_{0}^{\prime}(\kappa^{\prime})-\kappa^{2}j_{0}(\kappa^{\prime})j_{0}^{\prime}(\kappa),\label{eq:3.20}
\end{align}
and $\kappa=pR,$ $\kappa^{\prime}=p^{\prime}R.$

We expand
\begin{equation}
\kappa^{\prime}=\kappa+\eta-\frac{1}{2}\eta^{2}\,\frac{1}{\kappa}\label{eq:3.21}
\end{equation}
to $\mathcal{O}(\eta^{2}),$ noting than this series only converges
on $|\eta|<\frac{1}{2}pR,$ which is $|V_{0}|<E,$ where $|V_{0}|=|\lambda|/R$
is the height or depth of the potential and $E=p^{2}/2m$ is the energy
of the projectile. For $|\eta|>\frac{1}{2}pR,$ the solutions are
real exponentials on $0\leq r\leq R.$ For a small number of terms
of a perturbative expansion to give a good approximation requires
the further constraint $|\eta|\ll1.$ Expanding the expression for
$\exp(i2\delta_{0}(p))$ in Eq. (\ref{eq:3.19}) to $\mathcal{O}(\eta^{2})$
with Mathematica \citep{Mathematica2019} gives phase shifts in agreement
with Eqs. (\ref{eq:3.11}) and (\ref{eq:3.17}).

\section{\label{sec:The-Wronskian-formula}The Wronskian formula}

A useful formula regarding phase shifts was obtained using the Wronskian
theorem \citep{Messiah1961}. We consider the two second order differential
equations
\begin{equation}
\{-\frac{1}{2m}\frac{d^{2}}{dr^{2}}+\frac{l(l+1)}{2mr^{2}}+V(r)\}y_{l}^{(i)}(r,p)=\frac{p^{2}}{2m}y_{l}^{(i)}(r,p)\label{eq:4.1}
\end{equation}
 and
\begin{equation}
\{-\frac{1}{2m}\frac{d^{2}}{dr^{2}}+\frac{l(l+1)}{2mr^{2}}\}y_{l}^{(f)}(r,p)=\frac{p^{2}}{2m}y_{l}^{(f)}(r,p).\label{eq:4.2}
\end{equation}
The Wronskian theorem gives
\begin{align}
W(y_{l}^{(i)}(r,p),y_{l}^{(f)}(r,p))|_{0}^{M} & =\int_{0}^{M}dr\,\{y_{l}^{(i)}(r,p)y_{l}^{(f)\prime\prime}(r,p)-y_{l}^{(i)\prime\prime}(r,p)y_{l}^{(f)}(r,p)\}\label{eq:4.3}\\
 & =-2m\int_{0}^{M}dr\,y_{l}^{(i)}(r,p)V(r)y_{l}^{(f)}(r,p).\nonumber 
\end{align}
For potentials, $V(r),$ that fall off faster than $1/r$ as $r\rightarrow\infty$
and may diverge at the origin no faster than $1/r^{2},$ we know the
asymptotic behaviour
\begin{equation}
y_{l}^{(f)}(r,p)\rightarrow\sqrt{\frac{2}{\pi}}\,\sin(pr-l\frac{\pi}{2})\quad\mathrm{and}\quad y_{l}^{(i)}(r,p)\rightarrow\sqrt{\frac{2}{\pi}}\,\sin(pr-l\frac{\pi}{2}+\delta_{l}(p)).\label{eq:4.4}
\end{equation}
The Wronskian vanishes at the origin and, for $pM\gg1$ approaches
\begin{equation}
W(y_{l}^{(i)}(M,p),y_{l}^{(f)}(M,p))\rightarrow\frac{2p}{\pi}\,\sin(\delta_{l}(p)),\label{eq:4.5}
\end{equation}
so
\begin{equation}
\sin(\delta_{l}(p))=-\frac{\pi m}{p}\int_{0}^{\infty}dr\,y_{l}^{(i)}(r,p)V(r)y_{l}^{(f)}(r,p).\label{eq:4.6}
\end{equation}

We have Eq. (\ref{eq:2.16}) as our approximation of the interacting
wavefunction to first order in the potential. Inserting that gives,
to second order,
\begin{multline}
\sin(\delta_{l}^{(1)}(p)+\delta_{l}^{(2)}(p))=\delta_{l}^{(1)}(p)+\delta_{l}^{(2)}(p)\\
=-\frac{\pi m}{p}\int_{0}^{\infty}dr\,y_{l}^{(f)}(r,p)V(r)y_{l}^{(f)}(r,p)+\frac{\pi m}{p}\int_{0}^{\infty}dr\,\{\int_{0}^{\infty}dk\,y_{l}^{(0)}(r,k)\,\frac{2m\langle\,k,l;f\,|\,V\,|\,p,l;f\,\rangle}{k^{2}-p^{2}}\}V(r)y_{l}^{(f)}(r,p)\\
=-\frac{\pi m}{p}\langle\,p,l;f\,|\,V\,|\,p,l;f\,\rangle+\frac{2\pi m^{2}}{p}\,P\int_{0}^{\infty}dk\,\frac{\langle\,k,l;f\,|\,V\,|\,p,l;f\,\rangle^{2}}{k^{2}-p^{2}}.\label{eq:4.7}
\end{multline}
So the contributions to the phase shifts are
\begin{equation}
\delta_{l}^{(1)}(p)=-\frac{\pi m}{p}\langle\,p,l;f\,|\,V\,|\,p,l;f\,\rangle,\quad\delta_{l}^{(2)}(p)=\frac{2\pi m^{2}}{p}\,P\int_{0}^{\infty}dk\,\frac{\langle\,k,l;f\,|\,V\,|\,p,l;f\,\rangle^{2}}{k^{2}-p^{2}}.\label{eq:4.8}
\end{equation}

We verified that these two expressions give results in agreement with
Eqs. (\ref{eq:3.11}) and (\ref{eq:3.17}) for the spherical well
or barrier in the regime $pR\gg1.$

\section{\label{sec:Comparison-with-the}Comparison with the Green function
method}

We compare our method with the commonly used Green function method
\citep{Messiah1961} for solving the second order differential equation
that is the radial Schrödinger equation. This is an iterative perturbation
method. We take as the ``free'' Hamiltonian
\begin{equation}
H_{0}=-\frac{1}{2m}\frac{d^{2}}{dr^{2}}+\frac{l(l+1)}{2mr^{2}},\label{eq:5.1}
\end{equation}
with known fundamental solutions
\begin{equation}
\bar{j}_{l}(kr)=\sqrt{\frac{2}{\pi}}\,kr\,j_{l}(kr)\quad\mathrm{and}\quad\bar{n}_{l}(kr)=\sqrt{\frac{2}{\pi}}\,kr\,n_{l}(kr),\label{eq:5.2}
\end{equation}
the latter being singular at the origin for $l\geq1.$ So we want
to solve
\begin{equation}
\langle\,r\,|\,H_{0}-\frac{k^{2}}{2m}\,|\,k;i\,\rangle=-V(r)\langle\,r\,|\,k;i\,\rangle.\label{eq:5.3}
\end{equation}
We define the Green function, $G(r,r^{\prime}),$ as a solution of
\begin{equation}
\{-\frac{1}{2m}\frac{d^{2}}{dr^{2}}+\frac{l(l+1)}{2mr^{2}}-\frac{k^{2}}{2m}\}\,G(r,r^{\prime})=\delta(r-r^{\prime}).\label{eq:5.4}
\end{equation}
Then we note that the solutions, $\langle\,r\,|\,k;i\,\rangle,$ of
the integral equation
\begin{equation}
\langle\,r\,|\,k;i\,\rangle=\langle\,r\,|\,k;f\,\rangle-\int_{0}^{\infty}dr^{\prime}\,G(r,r^{\prime})\,V(r^{\prime})\,\langle\,r^{\prime}\,|\,k;i\,\rangle\label{eq:5.5}
\end{equation}
satisfy Eq. (\ref{eq:5.3}), the radial Schrödinger equation.

The integral equation is solved by iteration, first inserting the
free solution that vanishes at the origin, $\bar{j}_{l}(kr^{\prime}),$
in place of $\langle\,r^{\prime}\,|\,k;i\,\rangle$ on the right hand
side to generate $\langle\,r\,|\,k;i\,(1)\,\rangle$ to first order
in $\lambda.$ Next $\langle\,r\,|\,k;i\,(1)\,\rangle$ is inserted
on the right hand side to generate $\langle\,r\,|\,k;i\,(2)\,\rangle$
and so on.

For the Green function, there is freedom in the definition, and we
take the symmetric form
\begin{equation}
G(r,r^{\prime})=-\frac{\pi m}{k}\begin{cases}
\bar{j}_{l}(kr)\bar{n}_{l}(kr^{\prime}) & \mathrm{for}\ r<r^{\prime},\\
\bar{n}_{l}(kr)\bar{j}_{l}(kr^{\prime}) & \mathrm{for}\ r>r^{\prime}.
\end{cases}\label{eq:5.6}
\end{equation}

Unlike the method presented in this paper, the Green function method
is not designed to preserve the normalization of the solution. Using
the symmetric Green function guarantees that the solution has the
correct normalization to $\mathcal{O}(\lambda),$ but not at second
and higher order, as we have found. A process of renormalization is
necessary. If we find that the asymptotic form of the $n$-th order
solution as $r\rightarrow\infty$ is
\[
\langle\,r\,|\,k;i\,(n)\,\rangle=\sqrt{\frac{2}{\pi}}\,\sin(kr-l\frac{\pi}{2})\,\mathcal{A}_{l}^{(n)}+\sqrt{\frac{2}{\pi}}\,\cos(kr-l\frac{\pi}{2})\,\mathcal{B}_{l}^{(n)},
\]
then we must take
\begin{equation}
\cos(\Delta_{l}^{(n)}(k))=\frac{\mathcal{A}_{l}^{(n)}}{\sqrt{\mathcal{A}_{l}^{(n)2}+\mathcal{B}_{l}^{(n)2}}}\quad\mathrm{and}\quad\sin(\Delta_{l}^{(n)}(k))=\frac{\mathcal{B}_{l}^{(n)}}{\sqrt{\mathcal{A}_{l}^{(n)2}+\mathcal{B}_{l}^{(n)2}}},\label{eq:5.7}
\end{equation}
where $\Delta_{l}^{(n)}(k)$ is the total phase shift up to $n-$th
order,
\begin{equation}
\Delta_{l}^{(n)}(k)=\delta_{l}^{(1)}(k)+\dots+\delta_{l}^{(n)}(k).\label{eq:5.8}
\end{equation}

At first order, we encounter the integrals
\begin{align}
\mathcal{A}_{l}^{(1)}(r) & =\frac{\pi m}{k}\int_{0}^{r}dr^{\prime}\,\sqrt{\frac{2}{\pi}}\,kr^{\prime}\,n_{l}(kr^{\prime})\,V(r^{\prime})\,\sqrt{\frac{2}{\pi}}\,kr\,j_{l}(kr^{\prime}),\nonumber \\
\mathcal{B}_{l}^{(1)}(r) & =\frac{\pi m}{k}\int_{0}^{r}dr^{\prime}\,\sqrt{\frac{2}{\pi}}\,kr^{\prime}\,j_{l}(kr^{\prime})\,V(r^{\prime})\,\sqrt{\frac{2}{\pi}}\,kr^{\prime}\,j_{l}(kr^{\prime}).\label{eq:5.9}
\end{align}
We find
\begin{equation}
\delta_{l}^{(1)}(k)=-\mathcal{B}^{(1)}(\infty),\label{eq:5.10}
\end{equation}
in agreement with Eq. (\ref{eq:3.11}). If $r$ is increased without
bound to find $\mathcal{B}^{(1)}(\infty),$ the potential must fall
off faster than $1/r^{\prime}$ for convergence. Thus this method
cannot be applied to the Coulomb potential, as we found for our method.
At small $r,$
\begin{equation}
\mathcal{A}^{(1)}(r)\sim C\,\int_{0}^{r}dr^{\prime}\,\frac{1}{r^{\prime l}}V(r^{\prime})r^{\prime(l+1)},\label{eq:5.11}
\end{equation}
so $V(r^{\prime})$ may diverge at the origin provided the divergence
is slower than $1/r^{\prime2}.$

At second order, we encounter integrals such as
\begin{equation}
\mathcal{C}^{(2)}(\infty)=\frac{\pi^{2}m^{2}}{k^{2}}\int_{0}^{\infty}dr^{\prime}\,\bar{n}_{l}(kr^{\prime})\,V(r^{\prime})\,\bar{n}_{l}(kr^{\prime})\int_{0}^{r^{\prime}}dr^{\prime\prime}\,\bar{j}_{l}(kr^{\prime\prime})\,V(r^{\prime\prime})\,\bar{j}_{l}(kr^{\prime\prime}).\label{eq:5.12}
\end{equation}
This will converge if the bounds just found are satisfied. This is
a nested double integral, unlike what we found with the unitary method.

As a check on the validity of this method, we again considered the
spherical well/barrier for $s$-wave scattering ($l=0$). The integrals,
such as that in Eq. (\ref{eq:5.12}), were straightforward to evaluate
for this finite range potential. We expect they would pose more difficulty
in the general case. The simplicity of the unitary method, where all
integrals we encountered contained approximations to the delta function,
compared to evaluating such integrals, is clear. We found complete
agreement with the results of Eqs. (\ref{eq:3.11}) and (\ref{eq:3.17}).

\section{\label{sec:Conclusions}Conclusions}

The aim of this paper was to investigate the consequences of relating
an interacting Hamiltonian to the corresponding free Hamiltonian by
a unitary transformation. Using the example of scattering from a rotationally
invariant potential with the radial Schrödinger equation, we demonstrated
how to solve for the transformation to second order in the coupling
constant.

As a consequence of these results, we were able to formulate a perturbation
theory with significant differences from the commonly used Green function
method. The steps in the unitary method involve integrals over momentum
while those of the Green function method involve integrals over position.
We tested the unitary method on the spherical well and barrier, for
which exact solutions are known. Obtaining the $s-$wave phase shifts
in the regime where $pR\gg1$ ($p$ is the momentum under consideration
and $R$ is the finite range of the potential), we found agreement
with the Green function method and with the exact solution. It is
of note that the unitary method gives analytic approximations to the
wavefunction, while the Green function method gives piecewise continuous
approximations (with continuous first derivatives). The exact solution
has this character for this model.

The most significant difference between the unitary method and the
Green function method is that the former preserves wavefunction normalization
at all orders, while the latter requires renormalization.

The momentum integrals we encountered in the unitary method all contained
approximations to delta functions, greatly simplifying the calculation.
Of course away from the regime $pR\gg1$ the integrals would become
more involved. The Green function method position integrals are nested
at second order. This was not a problem for the finite range example,
but would add complication for a general, continuous, potential.

We comment on the Coulomb scattering problem, with solutions not accessible
by this unitary perturbation theory or with the Green function method
(which is known in this case as the Born approximation \citep{Born1926}).
Yet the exact solutions of the Coulomb problem are known. It should
be possible to perturb around those solutions, if the perturbing potential
is within the class for which perturbation theory is applicable.

So we consider the problem with Hamiltonian
\begin{equation}
H=H_{C}+V,\quad H_{C}=-\frac{1}{2m}\frac{d^{2}}{dr^{2}}+\frac{l(l+1)}{2mr^{2}}+\frac{Z_{1}Z_{2}\alpha}{r}.\label{eq:6.1}
\end{equation}
Our results carry over with $y_{l}^{(f)}(r,p)$ replaced by $y_{l}^{(C)}(r,p),$
the exact Coulomb solutions. Another Wronskian result can be derived
in this case. With the asymptotic forms \citep{Weinberg2015}
\begin{equation}
y_{l}^{(C)}(r,p)\rightarrow\sqrt{\frac{2}{\pi}}\,\sin(pr-l\frac{\pi}{2}-\eta_{C}\ln(2pr)+\sigma_{l}(p))\quad\mathrm{and}\quad y_{l}^{(i)}(r,p)\rightarrow\sqrt{\frac{2}{\pi}}\,\sin(pr-l\frac{\pi}{2}-\eta_{C}\ln(2pr)+\delta_{l}(p)),\label{eq:6.2}
\end{equation}
we have
\begin{equation}
\sin(\delta_{l}(p)-\sigma_{l}(p))=-\frac{\pi m}{p}\,\int_{0}^{\infty}dr\,y_{l}^{(i)}(r,p)V(r)y_{l}^{(C)}(r,p),\label{eq:6.3}
\end{equation}
where $\sigma_{l}(p)$ are the Coulomb phase shifts \citep{Weinberg2015}
and $\eta_{C}$ is defined in Eq. (\ref{eq:3.13}). Use of this formula
would give a result to all orders in $\eta_{C}$ and any desired order
in the perturbation.

\bibliographystyle{vancouver}

\end{document}